\documentclass[11pt]{article}
\usepackage{latexsym}
\usepackage{graphicx}

\input{tcilatex}

\begin{document}

\title{Division of Labor as the Result of Phase Transition}
\author{Jinshan Wu$^{1,2}${\thanks{%
Electrical address: wujinshan@yahoo.com}}, Zengru Di$^1$, Zhanru Yang$^{1,2}$
\\
1. Department of Physics, Beijing Normal University, Beijing, P.R.China \\
2. Institute of Theocratical Physics, Beijing Normal University, Beijing,
P.R.China}
\maketitle

\begin{abstract}
The emergence of labor division in multi-agent system is analyzed by the
method of statistical physics. Considering a system consists of N
homogeneous agents. Their behaviors are determined by the returns from their
production. Using the Metropolis method in statistical physics, which in
this model can been regarded as a kind of uncertainty in decision making, we
constructed a Master equation model to describe the evolution of the agents
distribution. When we introduce the mechanism of learning by doing to
describe the effect of technical progress and a formula for the competitive
cooperation, the model gives us the following interesting results: (1) As
the results of long term evolution, the system can reach a steady state. (2)
When the parameters exceed a critical point, the labor division emerges as
the result of phase transition. (3) Although the technical progress decides
whether or not phase transition occurs, the critical point is strongly
effected by the competitive cooperation. From the above physical model and
the corresponding results, we can get a more deeply understanding about the
labor division.
\end{abstract}


Pacs:05.45.-a; 87.23-n

\section{Introduction}

The economic system is no doubt a many-particle system-it can be viewed as a
collection of numerous interaction agents. So it is possible that methods
and concepts developed in the study of strongly fluctuation systems might
yield new results in this area. In fact, in the past decades the approaches
from statistical physics have been applied in economics and a lot of
interesting results, including empirical laws and theoretical models, have
been achieved (see \cite{Mantegna, Amaral, ponzi, plerou} and \cite{criss}
as a review of Econophysics). Among all these studies, a great deal of
researchers is on the agent-based modelling and related non-trivial
self-organizing phenomena. In the economic system, the agents learn from
each other, and their activities may be influenced by others' actions. These
interactions between agents may be simple and local, but they may have
important consequence related with the emergence of global structure. To
understand the mechanism behind these innovation phenomena, the methods and
concepts in phase transitions and critical phenomena are helpful. For
instance, in the study of majority and minority game\cite{challet1,
challet2, marsili}, opinion formation\cite{krzysztof}, and computational
ecosystems\cite{kephart}.

In this paper, we focus on the formation of labor division. Roughly
speaking, an economic organizational pattern is said to involve the division
of labor if it allocates the labor of different individuals to different
activities. Hence the specialization of individuals and the number of
professional activities are two sides of division of labor\cite{yang}. It is
a common functional organization observed in many complex systems, and it is
a fundamental way to improve efficiency and utilization so as to get global
optimization for the system. In order to investigate the mechanism behind
the formation of labor division, we have constructed a simple model with
many interacting agents. Every agent has only two kind of tasks, namely $A$
and $B.$ we describe the level of specialization of agent by his
working-time share spent on producing $A$ or $B$. Each agent make their
decisions for working-time in different tasks, and receive payoffs according
to their and other agents' choices. The agents can adapt by evaluating the
performance of their strategies from past experience so as to get maximum
returns. The returns for any agent is determined by its production with
endogenous technical progress-through the mechanism of learning by doing,
and its cooperation with other agents. Just like the Hamiltonian in
statistical physics, the payoff function determines the behavior of the
agent in economic system. Because of the bounded rationality and
incompleteness information in the system, we have introduced a parameter,
named social temperature $T$ (in the model, we have absorbed the $%
\beta=k_{B}T$ into the other parameters. Such an approach is traditionally
used in Statistical Physics, for example, let $H^{^{\prime}}=\beta H$, so
the new $J$ in Hamiltonian of Ising model means $\beta J$ actually.), to
describe the degree of randomness in decision-making. Then we assume that
the system should obey the canonical ensemble distribution, that is the
probability $P\left(\vec{x}\right) $ of a microstate $\vec{x}$ is
proportional to its `Boltzmann factor' determined by the total returns of
the microstate $\vec{x}$:

\begin{equation}
P\left( \vec{x}\right) =\frac{e^{+\beta E\left( \vec{x}\right) }}{\int
e^{+\beta E\left( \vec{x}\right) }d\vec{x}}  \label{jjfenbu1}
\end{equation}

Then using the Metropolis simulation method\cite{binder, zhu}, we can get a
Master equation to investigate the evolution of the system. With the
continual change of system parameters, we have found a so called `social
phase transitions' phenomenon related to the emergence of labor division.

The model is defined in Section 2. The economies of specialization is
introduced by increasing returns from learning by doing. And the economies
of complementarity is described by an additional payoffs from the
combination of two products. Section 3 gives the numerical results. The
effects of parameters on critical point are discussed in detail. It is
revealed that although the technical progress is a key factor that determine
whether or not the labor division will happen, the competitive cooperation
among agents has very important effects on the critical point. Our results
are summarized in Section 4.

\section{The Model}

Let's consider a system consists of $N$ homogeneous agents. In a given time
period $T$, any agent has two different kind of tasks, namely $A$ and $B$.
In any time unit, the fraction of working time on task $A$ for $ith$ agent
is denoted by $p^{i}\left( t\right) \in \left[ 0,1\right] $. we are
interested in the long-term ($T\rightarrow \infty $) evolution of the
system, especially the emerging of labor division.

$p^{i}$ is a real number in $\left[ 0,1\right] $, which describes the
working pattern for the $ith$ agent. If $p^{i}$ equals $1$ or $0$, the agent
is full specialized in $A$ or $B$ and we call the system is in complete
division of labor. If $p^{i}$ equals $0.5$, it means that the agent spends
same time interval on tasks $A$ and $B$, and we call it as the time-dividing
working mode in the following discussion. Usually $p^{i}$ could be any real
number between $0$ and $1$, in this case the agent does not have any
preference on job $A$ or $B.$We focus on the global behavior of the system.
Based on the above description of the agent, we introduce the following two
order parameters to describe the behavior of the system on macroscopic
level: 
\begin{equation}
\lambda _{1}\left( t\right) =\frac{1}{N}\sum_{i}\left| 2p^{i}\left( t\right)
-1\right|
\end{equation}%
\begin{equation}
\lambda _{2}\left( t\right) =\frac{1}{N}\sum_{i}\left( 2p^{i}\left( t\right)
-1\right)
\end{equation}%
Where $\lambda _{1}$ describes the intensity of labor division and
cooperation in the system. It has three special values, $\left\{
0,0.5,1\right\} $, represent time-dividing working mode, every $p^{i}$
equals $0.5$, no-preference working mode, $p^{i}$ distributed in (0,1)
randomly, and full specialization, every $p^{i}$ equals $0$ or $1$,
respectively. $\lambda _{2}$ gives the agents allocation on tasks $A$ and $B$
on global (in average on macroscopic level).

In the following discussion, we try to specify a suitable evolution rule for 
$p^{i}\left( t\right) $ based on the similar approach in statistical
physics. So we can then determine the dynamical equations for $p^{i}\left(
t\right) $, and get the final steady states for $\lambda _{1}\left( t\right) 
$ and $\lambda _{2}\left( t\right) $.

Analogous with Ising model in statistical physics, $p^{i}$ could be treated
as the spin in $ith$ point of the lattice, and then $\lambda _{2}$ is the
magnetization. So if we can give the Hamiltonian of the system, the
evolution of the system could be determined. In the economic system, the
agent switches his behavior according to his evaluations on the returns from
production. Every agent try to get maximum return just as any practical
tends to stay in the state with lowest energy. So the payoff function from
the agent's production should have the same effect as the Hamiltonian. The
working mode for each agent should be determined by its returns.

\subsection{Production function}

According to the previous literatures on specialization and economic
organization, we know that there are several factors that related to the
division of labor, such as the increasing returns to specialization and
transaction costs. In fact, limited ability on learning and incomplete
information on the technology will lead to the labor division directly. But
we don't take them into account in this model. In our model, every agent
know the technology for producing $A$ and $B$. And for any agent there is
not any comparative advantages for producing $A$ or $B$ in the initial. The
technical progress is achieved by the mechanism of learning by doing without
any cost. The payoff function for agent $i$ is given by the following
formulas: 
\begin{equation}
\varepsilon ^{i}=\varepsilon _{tech}^{i}+\varepsilon _{self}^{i}+\varepsilon
_{co}^{i}  \label{single}
\end{equation}%
where 
\begin{equation}
\varepsilon _{tech}^{i}=p^{i}\times \Gamma _{A}^{i}+\alpha \left(
1-p^{i}\right) \times \Gamma _{B}^{i}  \label{tech}
\end{equation}%
\begin{equation}
\varepsilon _{self}^{i}=\beta \min \left( p^{i}\times \Gamma _{A}^{i},\left(
1-p^{i}\right) \times \Gamma _{B}^{i}\right)  \label{self}
\end{equation}%
\begin{equation}
\varepsilon _{co}^{i}=\frac{1}{2}\beta \left( \Gamma _{B}^{i}\left(
1-p^{i}\right) -\Gamma _{A}^{i}p^{i}\right) \sum_{j\neq i}\left( \Gamma
_{A}^{j}p^{j}-\Gamma _{B}^{j}\left( 1-p^{j}\right) \right)  \label{Eco}
\end{equation}%
In the functions above, $\Gamma _{A}^{i}$ and $\Gamma _{B}^{i}$ are the
technology for producing $A$ and $B$ respectively. For simplicity and
without losing any generality, we assume that unit production $A$ will get
unit return. And $\alpha \in \left( 1,\infty \right) $ is the return of unit
production $B$. $\varepsilon _{self}^{i}$ gives the additional benefits for
composite $A$ and $B$ into a final product. So it is related to the less one
between $A$ and $B$. If agent $i$ has some more single product after the
composition himself, he can also get another return, named $\varepsilon
_{co}^{i}$, from the cooperation with other agents. But the return got from
the composition with the product from other agent is factored by $\frac{1}{2}%
\beta $. Because of the incomplete information in the system, any agent
couldn't know the situation of production on global. So in order to get the
corresponding returns $\varepsilon _{co}^{i}$ for every agent, we used the
average-field approach similar with that in statistical physics. That means,
for every agent, he will match his product with all the other agents and
then the result is the average. This assumption indicate that there are
already a public market in economics. In the function for $\varepsilon
_{co}^{i}$, $\left( \Gamma _{B}^{i}\left( 1-p^{i}\right) -\Gamma
_{A}^{i}p^{i}\right) $ gives the surplus product $B$ for agent $i$, and the
sum of $\left( \Gamma _{A}^{j}p^{j}-\Gamma _{B}^{j}\left( 1-p^{j}\right)
\right) $ gives the surplus product $A$ of the all other agents. When these
two terms have different signs, their product is negative and $\varepsilon
_{co}^{i}$, the corresponding payoff from the cooperation, will be also
negative. In this case, the surplus product of the agent is the same kind as
the final surplus of all the other agents. Because of the transaction cost
and the diseconomies of incomplementarity, it is rationale that the agent
would get negative payoff.

\subsection{Two descriptions for learning by doing}

The way for technical progress is learning by doing. This mechanism is
related to the accumulation of agent's historical production behavior. That
means the development of comparative advantages is determined by related
working time. Let $x_{AorB}^{i}$ denote the integral working time on $A$ or $%
B$ for agent $i$:

\begin{equation}
\begin{array}{c}
x_{A}^{i}=\int_{0}^{t}d\tau p^{i}\left( \tau \right) \\ 
x_{B}^{i}=\int_{0}^{t}d\tau (1-p^{i}\left( \tau \right) )%
\end{array}%
\end{equation}

\begin{figure}[tbp]
\centering \includegraphics{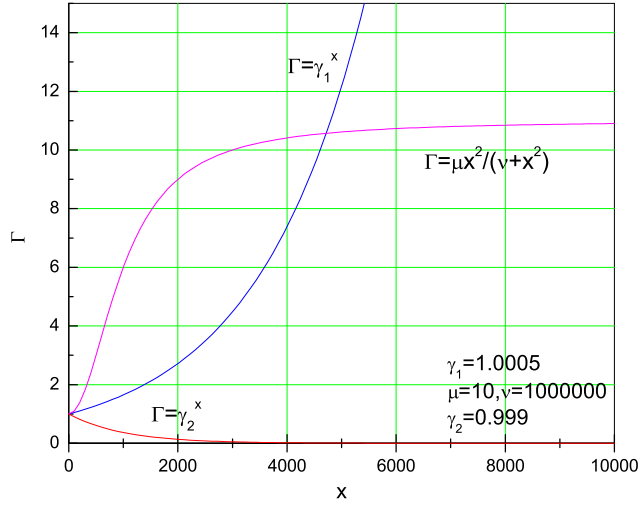}
\caption{Sample functions of $\Gamma ^{i}$ used in the simulations.
Parameters are labelled in the graph.}
\label{g_m}
\end{figure}

We have introduced two mechanisms for technical progress, named $\gamma $%
-mechanism and $\mu $-mechanism. They are given by following functions: 
\begin{equation}
\Gamma _{A/B}^{i}=\gamma ^{x_{A/B}^{i}}  \label{gamma}
\end{equation}

\begin{equation}
\Gamma _{A/B}^{i}=1+\frac{\mu {x_{A/B}^{i}}^{2}}{\nu +{x_{A/B}^{i}}^{2}}
\label{mu}
\end{equation}%
These two mechanisms will give different results. Equation(\ref{gamma}) is
an exponential function which gives unlimited growth on technology, while
there is an upper limit in equation(\ref{mu}). Figure(\ref{g_m}) shows three
functions for $\Gamma _{A/B}^{i}$ in our computer simulations.

\subsection{The Master equation for the system evolution}

In the production function discussed above, term $\varepsilon _{co}^{i}$ is
the return from the agents' cooperation. It introduces interactions among
all agents (the same as the interactions among spins). So if we discuss the
dynamics in $\mu $-space, we should construct $N$ evolutionary equations for 
$p^{i}$. But they are difficult for both theoretical analysis and computer
simulation. In the following discussion, we try to describe the evolution of
the system in $\Gamma $-space. That is to discuss the evolutionary behavior
of the joint density function $P\left( \left\{ p^{i}\right\} ,t\right) $ in
the space sponsored by $\{p^{i}\}$. Then the total production for the system
is: 
\begin{equation}
E=\sum_{i=1}^{N}\varepsilon _{i}  \label{total}
\end{equation}%
Where $\varepsilon _{i}$ is the production of agent $i$ given by eq(\ref%
{single}). The same as the approach of critical dynamics in Ising model, the
Master equation for the joint density function $P\left( \left\{
p^{i}\right\} ,t\right) $ in $\Gamma $ space is 
\begin{equation}
\frac{d}{dt}P\left( p^{1},\dots ,p^{i},\dots ,p^{N};t\right)
=\sum_{i}\int_{0}^{1}dx^{i}%
\begin{array}{c}
\left[ \omega ^{i}\left( x^{i}\rightarrow p^{i}\right) P\left( p^{1},\dots
,x^{i},\dots ,p^{N};t\right) \right. \\ 
\left. -\omega ^{i}\left( p^{i}\rightarrow x^{i}\right) P\left( p^{1},\dots
,p^{i},\dots ,p^{N};t\right) \right]%
\end{array}
\label{master}
\end{equation}%
where $\omega ^{i}$ is the transition probability which is determined by the
Boltzman factors as the following: 
\begin{equation}
\omega ^{i}\left( p^{i}\rightarrow x^{i}\right) =\frac{e^{\Delta E\left(
p^{i}\rightarrow x^{i}\right) }}{\int_{0}^{1}dx^{i}e^{\Delta E\left(
p^{i}\rightarrow x^{i}\right) }}  \label{jump}
\end{equation}%
In the above transition probabilities, $\Delta E\left( p^{i}\rightarrow
x^{i}\right) $ is the change of returns related to the variation of working
mode for $ith$ agent. From eq.(\ref{total}) we can get 
\begin{equation}
\Delta E\left( p^{i}\rightarrow x^{i}\right) =%
\begin{array}{c}
\left[ \left( \Gamma _{A}^{i}-\alpha \Gamma _{B}^{i}+\frac{1}{2}\beta \left(
\Gamma _{A}^{i}-\Gamma _{B}^{i}\right) \right) \left( x^{i}-p^{i}\right)
\right. \\ 
\left. -\frac{\beta }{2}\left( \left| x^{i}\Gamma _{A}^{i}-\left(
1-x^{i}\right) \Gamma _{B}^{i}\right| -\left| p^{i}\Gamma _{A}^{i}-\left(
1-p^{i}\right) \Gamma _{B}^{i}\right| \right) \right. \\ 
\left. +2\varepsilon _{co}^{i}(x^{i})-2\varepsilon _{co}^{i}(p^{i})\right]%
\end{array}%
\end{equation}%
With the master equation(\ref{master}) and certain initial conditions, we
can get evolving behaviors of $\lambda _{1}$ and $\lambda _{2}$. Then we can
discuss the phenomena on labor division. It is difficult to have some
theoretical results. So in the next section we will show some numerical
simulation results by Monte Carlo simulation method. Based on the Master
equation(\ref{master}) and corresponding transition probability(\ref{jump}),
the simulation is proceed on the following Metropolis algorithm:

\begin{enumerate}
\item For any given state with $p^{i}$ for agent $i$, a new state $x^{i}$ is
randomly selected;

\item If $\Delta E\left( p^{i}\rightarrow x^{i}\right) >0$, then the
transition from $p^{i}$ to $x^{i}$ is proceeded;

\item If $\Delta E\left( p^{i}\rightarrow x^{i}\right) \leq 0$, then a
random number $\xi \in \left( 0,1\right) $ is selected. If $\xi \leq
e^{\beta \Delta E}$, the transition from $p^{i}$ to $x^{i}$ is proceeded or
else the agent $i$ keeps the original state.

\item For another agent $j$, goto step 1.
\end{enumerate}

\begin{figure}[tbp]
\centering \includegraphics{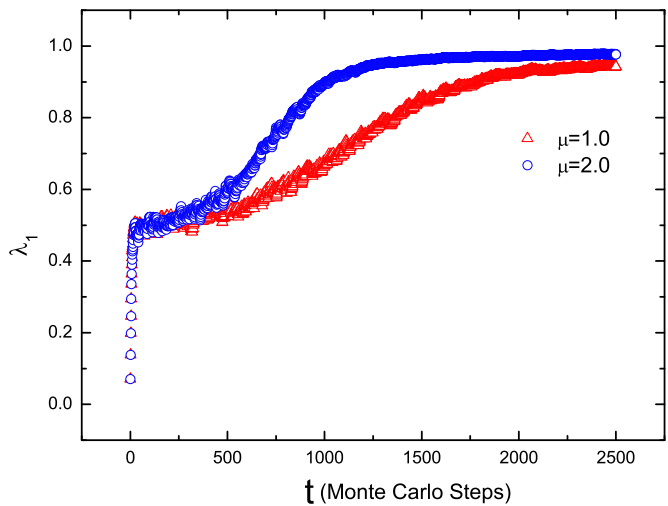}
\caption{A typical evolutionary behavior of the model under $\protect\mu$%
-mechanism. $\protect\beta =10.0$, $\protect\alpha =1.0$.}
\label{lambda-t}
\end{figure}

For any given initial state, the system will achieve a certain steady state
after some transient process. Figure(\ref{lambda-t}) gives a typical
evolution behavior under $\mu$ mechanism.

\section{The Results of Monte Carlo Simulation}

There are several parameters related to the mechanism of labor division in
the model. We will show the simple or maybe trivial results of $\gamma $%
-mechanism first and then emphasize our discussion on the results of $\mu $%
-mechanism. And we let $\alpha =1$ in all the simulations before we discuss
the effects of parameters $\alpha $ and $N$ on the system evolution in the
end.

\subsection{$\protect\gamma $-mechanism}

\begin{figure}[tbp]
\centering \includegraphics{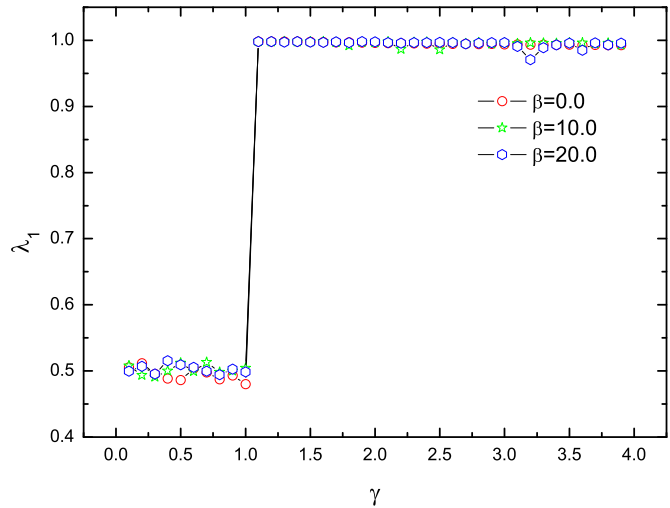}
\caption{Parameter $\protect\lambda_{1}$ in the final stationary state as a
function of parameter $\protect\gamma$ under $\protect\gamma$-mechanism. The
final steady state of the system is determined by $\protect\gamma>1$ or $%
\protect\gamma<1$.}
\label{ggamma}
\end{figure}

As shown in(\ref{ggamma}) because of $\gamma $-mechanism is described by an
exponential function, the system will achieve the state of labor division
for any given initial state and $\gamma >1$ no matter how small is $\gamma
-1 $. That means if the intensity of technical progress is big enough, the
agent would definitely specialized in a kind of task. Other factors such as
competitive cooperation described by parameter $\beta $ have no effects on
the long term evolution. 

\subsection{$\protect\mu$-mechanism}

\begin{figure}[tbp]
\centering \includegraphics{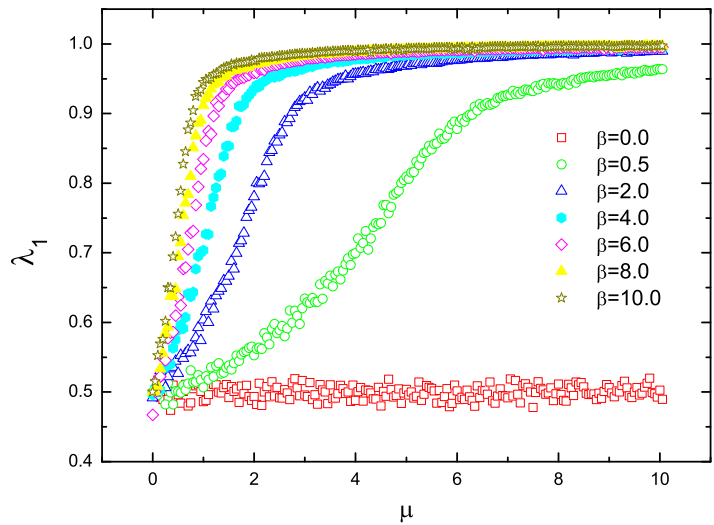}
\caption{For different parameter $\protect\beta$, parameter $\protect\lambda%
_{1}$ in the final stationary state as a function of parameter $\protect\mu$
under $\protect\mu$-mechanism. There is phase transition in the system. $%
\protect\beta$ has strongly effects on the critical point.}
\label{betamu}
\end{figure}

But for the model with $\mu $-mechanism, $\beta$ has some important effects
on the system evolution. As shown in Figure(\ref{betamu}), there are phase
transitions results in labor division and $\beta$ has strong effects on the
critical point $\mu_{c}$ for parameter $\mu$. $\mu _{c}$ decreases as the $%
\beta$ increasing. The $\mu$-mechanism for learning by doing gives a
logistic growth for technology. That is much more realistic than the $\gamma$%
-mechanism. The results reveal that the competitive cooperation among agents
is very important for the emergence of labor division. However, as discussed
in the end of this section, when $\mu =0$, that is there is no mechanism of
increasing returns, the system has no phase transition to labor division.
This result is rationale for the model and it is consistent with the theory
on specialization. The comparative advantages in production can only be
introduced by a positive feedback caused by increasing returns. So the
mechanism of learning by doing is a dominant factor for labor division. 

\begin{figure}[tbp]
\centering \includegraphics{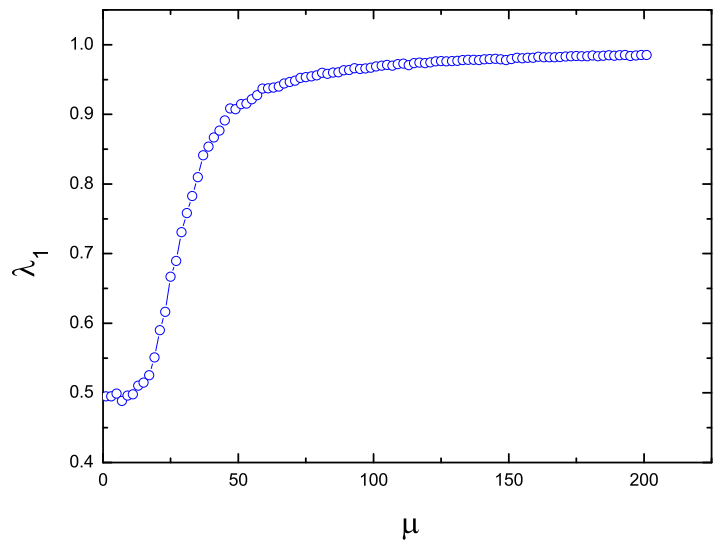}
\caption{Under $\protect\mu$-mechanism and without the term $\protect%
\varepsilon_{co}^{i}$ that is the payoff from the cooperation among agents,
parameter $\protect\lambda_{1}$ in the final stationary state as a function
of parameter $\protect\mu$ for different parameter $\protect\beta$. The
critical point $\protect\mu_{c}$ is much larger than that in Fig. 4 and even
in Fig. 6.}
\label{no_beta}
\end{figure}

In order to investigate the effects of all terms in production function in
detail, we have simulated some special cases. The first case is $\beta =0$.
Then $\varepsilon _{self}^{i}$ and $\varepsilon _{co}^{i}$ in equation (\ref%
{single}) are all zero and technical progress is the only one factor that
affects the final results. As shown in figure(\ref{no_beta}), there is a
phase transition in the system. Under the given conditions, the critical
point for $\mu$ ($\mu_{c}$) is around 50. 

\begin{figure}[tbp]
\centering \includegraphics{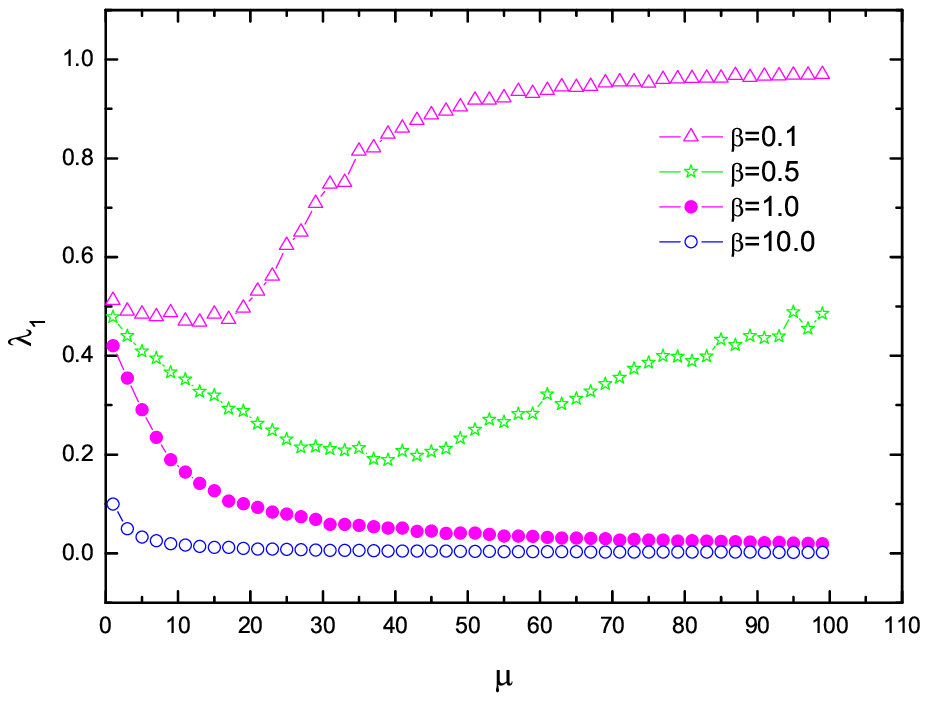}
\caption{Under $\protect\mu$-mechanism, parameter $\protect\lambda_{1}$ in
the final stationary state as a function of parameter $\protect\mu$ when
there is only the term for technical progress in the production function.
The critical point $\protect\mu_{c}$ is much larger than that in Fig. 4.}
\label{no_eco}
\end{figure}

Then we have simulated the model with $\varepsilon _{self}^{i}$ but no $%
\varepsilon _{co}^{i}$. The results are shown in (\ref{no_eco}). When $\beta 
$ is greater than a critical value $\beta _{c}$, the system will stabilized
in the state with $\lambda _{1}=0$. That is $p^{i}=0.5$ for any agent $i$.
Every agent does the tasks $A$ and $B$ himself and spends the same time on
both tasks. The system is in the time-dividing working mode. Only when $%
\beta $ is smaller than $\beta _{c}$, does the system have phase transition
to labor division. But the critical value $\mu _{c}$ is much larger than
that of in Figure (\ref{betamu}). It is even larger than the $\mu _{c}$ in
Figure (\ref{no_beta}). So the term $\varepsilon _{self}^{i}$ has the effect
of anti-specialization. From the simulation results in Figure (\ref{betamu})
and just as we have indicated in the beginning of this section, $\varepsilon
_{co}^{i}$ is helpful to labor division. Actually the term $\varepsilon
_{co}^{i}$ is the benefit for agent $i$ get from the cooperation with all
the other agents. It should has the same effect as the decreasing of
transaction costs. So it is not surprising that this term will enforce
specialization. 

\begin{figure}[tbp]
\centering \includegraphics{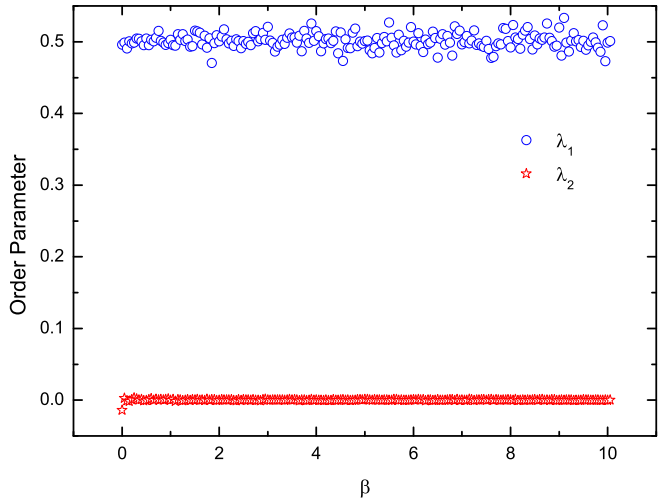}
\caption{Order parameters $lambda_{1}$ and $\protect\lambda_{2}$ as a
function of parameter $\protect\beta$ when there is no progress of
productivity. The system has no phase transition.}
\label{no_gamma}
\end{figure}

Another interesting case is the system without technical progress, that is $%
\Gamma _{A}^{i}$ and $\Gamma _{B}^{i}$ are all equal 1. In this case, the
system is exactly the same as Ising model. Let $S^{i}=2p^{i}-1$ and $\alpha
=1$, then the total returns for the system is

\begin{equation}
E=-\beta \sum_{i,j,i\neq j}S^{i}S^{j}+\frac{1}{2}\sum_{i}\beta \left(
1-\left| S^{i}\right| \right)  \label{Energy}
\end{equation}%
Where $S^{i}\in \left( -1,1\right) $. The distribution of the state for the
system is determined by $e^{E}$. Comprising with Ising model, let the
corresponding Hamiltonian is 
\begin{equation}
H=-\frac{1}{2}\sum_{i}\beta \left( 1-\left| S^{i}\right| \right) +\beta
\sum_{i\neq j}S^{i}S^{j}  \label{H}
\end{equation}%
and let $k_{\beta }T=1$, then the system could be described by Canonical
Ensemble with $e^{-\frac{H}{k_{\beta }T}}$. That is an anti-ferromagnetic
Ising model with global interaction. The similar approach shows that this
system either does't have phase transition related to labor division. The
result is shown in Figure(\ref{no_gamma}). 

\subsection{The effects of parameter $\protect\alpha$ and $N$}

\begin{figure}[tbp]
\centering \includegraphics{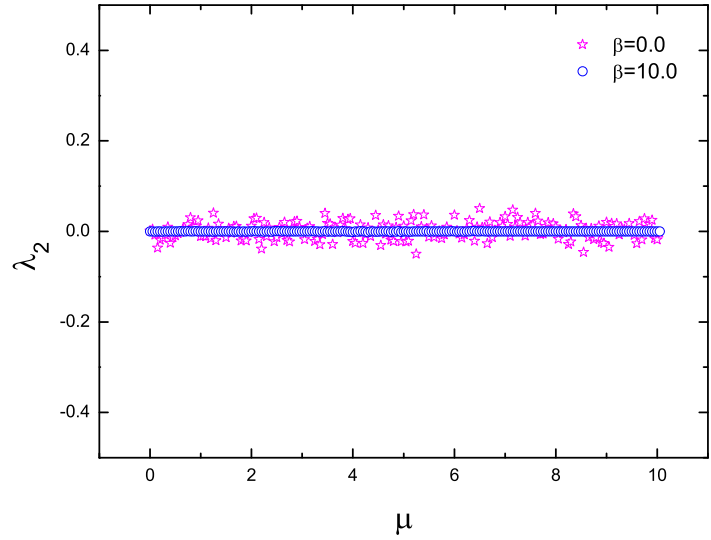}
\caption{Parameter $\protect\lambda_{2}$ in the final stationary state as a
function of parameter $\protect\mu$ for different $\protect\beta$ when $%
\protect\alpha=1$. There is no symmetry breaking with the phase transition.}
\label{lambda_2}
\end{figure}

\begin{figure}[tbp]
\centering \includegraphics{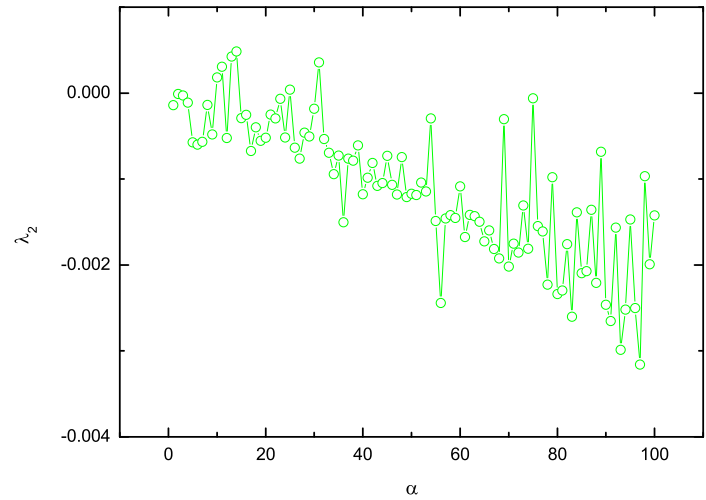}
\caption{Parameter $\protect\lambda_{2}$ in the final stationary state as a
function of parameter $\protect\alpha$. Because of the cooperation mechanism
described by parameter $\protect\beta$, $\protect\alpha$ has only a little
effect on the final distribution.}
\label{lambda_a}
\end{figure}

\begin{figure}[tbp]
\centering \includegraphics{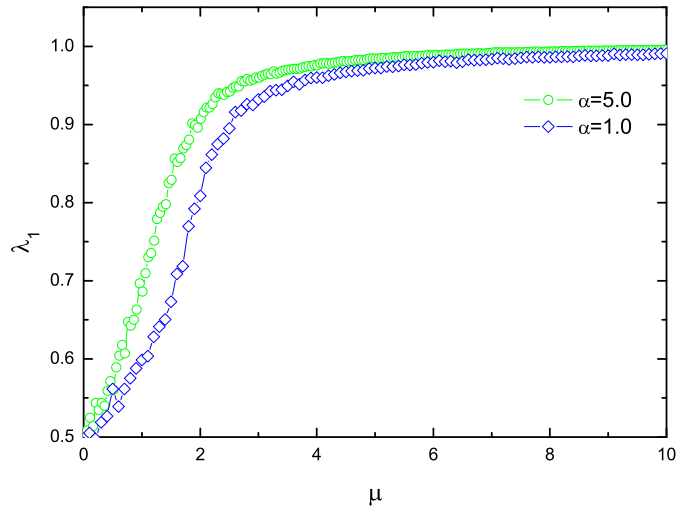}
\caption{Parameter $\protect\lambda_{1}$ in the final stationary state as a
function of parameter $\protect\mu$ for different $\protect\alpha$.}
\label{alpha5_1}
\end{figure}

Parameter $\lambda_{2}$ reflects the preference for tasks $A$ and $B$. So it
is expected that parameter $\alpha $, the ratio of returns for $A$ and $B$,
will affect final results on $\lambda_{2}$. The diagram in phase space for $%
\lambda_{2}$ when $\alpha=1$ is shown in Figure (\ref{lambda_2}), in which
we can see $\lambda_2$ fluctuates around zero, and no phase transition with
symmetry breaking emerges. And as shown in Figure(\ref{lambda_a}), $\alpha$
indeed has some effects on $\lambda_{2}$. But because of the combination
mechanism described by parameter $\beta$, $\alpha$ has only a little effect
on $\lambda_{2}$. It also has some detailed effects on labor division
described by $\lambda_{1}$, as shown in Figure (\ref{alpha5_1}). But $\alpha$
could not change the qualitative behavior of phase transition.

\begin{figure}[tbp]
\centering \includegraphics{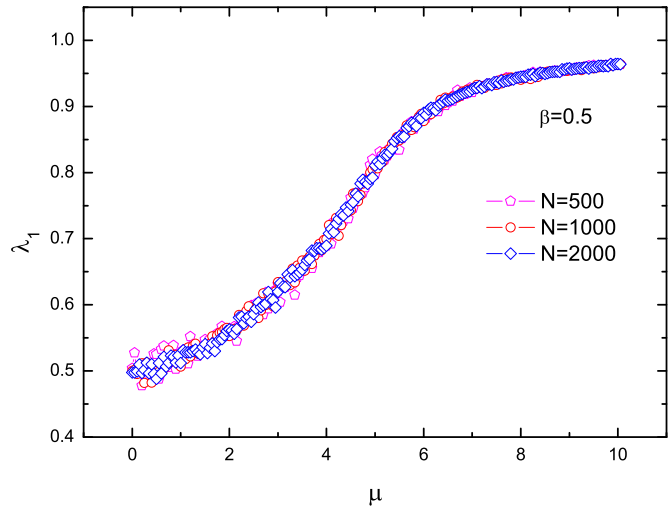}
\caption{When $\protect\beta=0.5,$ $\protect\alpha=1.0$, $\protect\lambda%
_{1}-\protect\mu$ curves for different population size $N$. The curves are
almost consistent.}
\label{500-1000}
\end{figure}

Here are other two phenomena that should be pay attention to, first, the
model has no scale effect for limited $N$, as shown in Figure (\ref{500-1000}%
). The phase curves for different $N$ are almost consistent. It has only
some effects on the accurate of simulation. Second, When $\gamma<1.0$ in $%
\gamma$-mechanism for technical progress, the productivity declines with the
time evolution. Then the system could not reach the state of specialization.
This result is also meaningful for real economic system. 

\section{Concluding Remarks}

A distinctive feature of the organization of a human society is the division
of labor. From the classical economic theory[11], the division of labor
comes from the development of endogenous comparative advantages. So there is
the intrinsic relationship between technical progress and the evolution of
the division of labor. But how and why it emerges from the system consists
of identical individuals? We have studied the formation of labor division by
the approach of statistical physics. The results reveal that there is a
phase transition with this pattern formation. Although the progress of
productivity dominated the phase transition occurs or not, the competitive
cooperation among the agents has important effects on the critical point. So
the market formation and labor division are usually reinforced each other.
All the above results give us deep understanding to the evolution of labor
division.

Studying the economy as an evolving complex system, we can avoid the
standard economic assumptions of equilibrium based on rational behavior by
agents. And the concepts and methods developed in statistical physics is
helpful to uncover fundamental principles governing the evolution of complex
adaptive systems. So the approaches presented here have potential
applications in a variety of economical, biological and financial problems.

\section{Acknowledgement}

This work was supported by the National Science Foundation under Grant
No.10175008 and National 973 Program under Grant No.G2000077307. We also
thanks to all members of Professor Yang's Group and faculties from
Department of System Science for their warm discussion.


\begin{thebibliography}{99}
\bibitem{Mantegna} R.N. Mantegna, H.E. Stanley, Scaling behavior in the
dynamics of an economic index, Nature 376 (1995) 46--49.

\bibitem{Amaral} LA,N, Amaral, P. Gopikrishnan, V. Plerou, H.E.Stanly, A
model for the growth dynamics of economic organizations, Physica A 299(2001)
127-136.

\bibitem{ponzi} A. Ponzi, Y. Aizawa, Evolutionary financial market models,
Physica A 287(2000) 507-523.

\bibitem{plerou} V. Plerou, P. Gopikrishnan, L.A.N. Amaral, M. Meyer, H.E.
Stanley, Scaling of the distribution of price fluctuations of individual
companies, Phys. Rev. E 60 (1999) 6519--6529.

\bibitem{criss} N.A. Criss, Review of Mantegna and Stanley, an introduction
to econophysics, Physics Today53(2000)12.

\bibitem{challet1} D. Challet, Y.-C. Zhang, Emergence of cooperation and
organization in an evolutionary game, Physica A 246(1997) 407.

\bibitem{challet2} D. Challet, M. Marsili, Phase transition and symmetry
braking in the minority game, Phys. Rev. E 60 (1999)6271.

\bibitem{marsili} M. Marsili, Market mechanism and expactations in minority
and majority games, Physica A 299(2001) 93-103.

\bibitem{krzysztof} Krzysztof Kacperski, Janusz A. Holyst Phase transitions
as a persistent feature of groups with leaders in models of opinion
formation, Physica A 287 (2000) 631-643.

\bibitem{kephart} J. O. Kephart, T. Hagg, and B. A. Huberman, Dynamics of
computational ecosystems, Phys. Rev. A 40, 404-421(1989); Collective
behavior of predictive agents, Physica D 106, 48-65(1990).

\bibitem{yang} X. Yang, Y.-K. Ng, Specialization and economic organization-a
new classical microeconomic framwork, North-Holland, 1993.

\bibitem{binder} K. Binder, D.W. Heermann, Monte Calo Simulation in
Statistacal Physics, Springer-Verlag, 1988.

\bibitem{zhu} Jian-Yang Zhu, Z. R. Yang, Exact solution of the kinetic
gaussian model, Phys. Rew. E59, (1999)1551.

\bibitem{arrow} K. J. Arrow, The economic implications of learning by doing,
Review of Economic Studies 29, 155-173(1962)
\end{thebibliography}
\end{document}